\documentclass{article}
\usepackage{fortschritte}
\usepackage{amssymb}
\usepackage{epsfig}
\usepackage{subfigure}
\def\beq{\begin{equation}}                     %
\def\eeq{\end{equation}}                       %
\def\bea{\begin{eqnarray}}                     
\def\eea{\end{eqnarray}}                       
\newcommand{\N}{{\kern+.25em\sf{N}\kern-.78em\sf{I} \kern+.78em\kern-.25em}}
\begin {document}                 
\def\titleline{
%
%
%
Non--Commutative Field Theories beyond Perturbation Theory
}
\def\email_speaker{
{\tt 
%
%
hofheinz@physik.hu-berlin.de
}}
\def\authors{
%
%
%
%
%
W.\ Bietenholz \1ad , F.\ Hofheinz \1ad \2ad          
and J.\ Nishimura \3ad\sp                           
}
\def\addresses{
%
%
%
%
\1ad
Humboldt Universit\"at zu Berlin, Invalidenstr. 110, 
D--10115 Berlin, Germany\\

\2ad                                        %
Freie Universit\"at Berlin, Arnimallee 14, D--14195 Berlin, Germany\\

\3ad                                        
Dept. Physics, Nagoya University, Nagoya 464--8602, Japan
}
\def\abstracttext{
%
%
%
We investigate two models in non-commutative (NC) field theory
by means of Monte Carlo simulations. Even if we start from the
Euclidean lattice formulation, such simulations are only feasible
after mapping the systems onto dimensionally reduced matrix models.
Using this technique, we measure Wilson loops in 2d NC gauge
theory of rank 1. It turns out that they are non-perturbatively 
renormalizable, and the phase follows an Aharonov-Bohm effect 
if we identify $\theta = 1/B$. Next we study the 3d $\lambda \phi^{4}$
model with two NC coordinates, where we present new results
for the correlators and the dispersion relation. We further reveal
the explicit phase diagram. The ordered regime splits 
into a uniform and a striped phase, as it was qualitatively
conjectured before. We also confirm the recent observation
by Ambj\o rn and Catterall that such stripes occur even in $d=2$,
although they imply the spontaneous breaking of translation symmetry.
However, in $d=3$ and $d=2$ we observe only patterns of
two stripes to be stable in the range of parameters investigated.
%
}
\large
\makefront

\vspace*{-10mm}

\section{NC field theories as matrix models}

NC field theory is highly fashionable, in particular due to its 
potential r\^{o}le as a low energy effective description of string 
and M theory. Here we study Euclidean NC field theory as such;
for a discussion of the Wick rotation and the related issue
of unitarity, see e.g.\ Ref.\ \cite{mink}.
We assume two coordinates to obey the standard NC relation
\beq  \label{standardNC}
[ \hat x_{\mu},\hat x_{\nu}] = i \theta \epsilon_{\mu \nu} \ .
\eeq
This implies an uncertainty relation 
$\Delta x_{\mu} \Delta x_{\nu} = O(\theta )$, and therefore UV 
and IR divergences are mixed. That property complicates perturbative
renormalization drastically \cite{MRS}. We can also observe 
non-perturbative manifestations of UV/IR mixing, see below. \\

{\bf NC gauge theory} can be formulated by a consistent use
of the star product.  Even for $U(1)$ a YM term appears in the
gauge action,
\beq
S[A] = \frac{1}{4} \int d^{d}x \, {\rm Tr} (F_{\mu \nu} \star
F_{\mu \nu}) \ , \quad 
F_{\mu \nu} = \partial_{\mu} A_{\nu} - \partial_{\nu} A_{\mu}
+ ig (A_{\mu} \star A_{\nu} - A_{\nu} \star A_{\mu}) .
\eeq
A lattice formulation can be written down, but its use in
simulations is hardly possible: already the generation of star gauge 
invariant link variables appears as a fatal obstacle.

However, $U(n)$ NC gauge theories can be mapped \cite{IIKK}
onto certain forms of the twisted Eguchi-Kawai model (TEK) \cite{TEK}
(which is defined on a single site),
\begin{displaymath}
S_{\rm TEK}[U] = -N \beta \sum_{\mu \neq \nu} Z_{\mu \nu}
{\rm Tr} (U_{\mu} U_{\nu} U_{\mu}^{\dagger} U_{\nu}^{\dagger}) \, ,
\ Z_{\mu \nu} = Z^{*}_{\nu \mu} = e^{2\pi i k /L} \
(\mu < \nu ,\ k \in \mathbb{N} ,\ L = N^{2/d}) ,
\end{displaymath}
where $U_{\mu}$ are unitary $N \times N$ matrices and $Z_{\mu \nu}$
is the twist. This mapping is based on ``Morita equivalence'', which
is exact at $N \to \infty$. At finite $N$ there is an exact mapping
onto a NC gauge theory on a lattice with spacing $a$ in a finite volume
\cite{AMNS}, if we choose $k = (N+1)/2$, which implies
$\theta = N a^{2}/\pi$. After this mapping the system can indeed be 
simulated \cite{2dU1}. 
Thus we by-pass the problems with a direct lattice formulation.
The limit $N\to \infty$
corresponds to the simultaneous continuum and infinite volume limit
(these limits are entangled due to UV/IR mixing).\\

The action of the {\bf NC $\lambda \phi^{4}$ model} reads
\beq
S[\phi ] = \int d^{d}x \, \Big[ \frac{1}{2} \partial_{\mu} \phi 
\partial_{\mu} \phi + \frac{m^{2}}{2} \phi^{2} + \frac{\lambda}{4}
\phi \star \phi \star \phi \star \phi \Big] \ .
\eeq
Here the star product does not affect the bilinear terms, so the
strength of the self-coupling $\lambda$ also determines the extent 
of NC effects.

We consider $d=3$ with a commutative time direction and two space 
coordinates obeying relation (\ref{standardNC}) \cite{Lat01}. 
\footnote{Including also the time in the NC relation seems
more consistent from the gravity inspired motivation for
NC geometry \cite{gravi}, but it causes especially severe problems 
related to causality \cite{SST}.}
Again we start
from a lattice formulation --- $ \phi(t, \vec x )$ is defined on a
$T \times N^{2}$ lattice --- and map it on a matrix model \cite{AMNS} 
\footnote{A corresponding matrix model formulation for 3d
fermions was suggested in Ref.\ \cite{mafia}.}.
Its action takes the form
\begin{displaymath}
S [ \hat \phi ] = {\rm Tr} \sum_{t=1}^{T} \biggl[
  \frac{1}{2}\sum_\mu\left(\Gamma_\mu\hat{\phi}(t)\,
  \Gamma_\mu^\dagger-\hat{\phi}(t)\right)^2\\
  +\frac{1}{2}\left(\hat{\phi}(t+1)-\hat{\phi}(t)\right)^2+
  \frac{m^2}{2}\hat{\phi}^2(t)+\frac{\lambda}{4}\hat{\phi}^4(t)
  \biggl]\ , \nonumber
\end{displaymath}
where $\hat \phi (t)$ are Hermitian $N\times N$ matrices.
The ``twist eaters'' $\Gamma_{\mu}$ provide the spatial shift,
since they obey the Weyl-'t Hooft commutation relation
$\Gamma_{\mu} \Gamma_{\nu} = Z_{\nu \mu} \Gamma_{\nu} \Gamma_{\mu}$
(where $Z$ is the same twist as in the TEK describing NC gauge theory).
The final formulation has a remarkable similarity to the 
corresponding model on a ``fuzzy sphere'' \cite{fuzzy}.

\section{Numeric results}

\subsection{Gauge theory on a NC plane}

We simulated pure 2d NC $U(1)$ gauge theory \cite{2dU1} --- for analytical 
work on this model, see Ref.\ \cite{PanSza}. The large $N$ limit at fixed
$\beta $ coincides with the planar limit of the commutative theory
\cite{TEK}, which was solved by Gross and Witten \cite{GW}.
It exhibits an exact area law for the Wilson loop. From this connection
we identify the ``physical area'' of one lattice plaquette as
$a^{2} \simeq 1/\beta$. However, in contrast to the planar limit
we take the {\em double scaling limit} where $N \to \infty$,
$\beta \to \infty$, so that $\theta \propto N/\beta$ is kept constant.
Thus we can extrapolate to a NC continuum theory in an infinite volume.

\begin{figure}[htbp]
 \begin{center} 
   \includegraphics[width=0.455\linewidth]{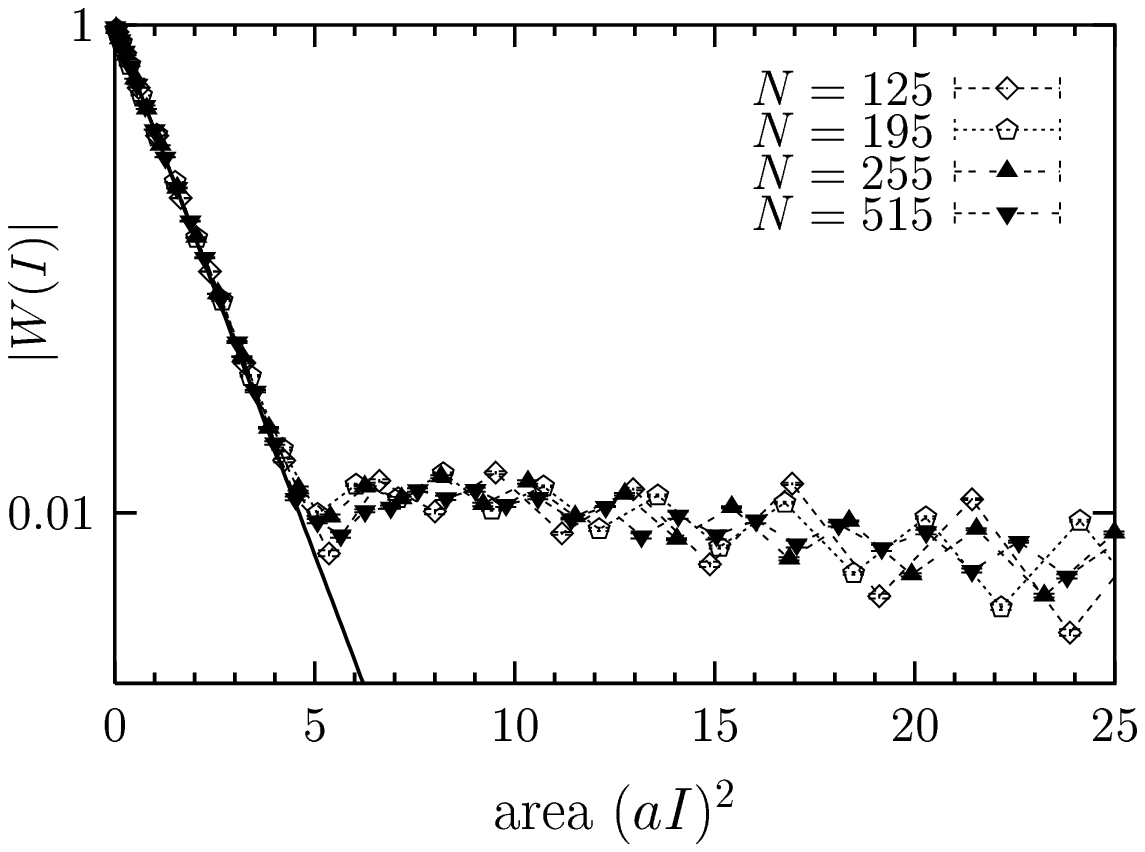}%
   \vspace{2cm}\hspace{.5cm}\includegraphics[width=.48
\linewidth]{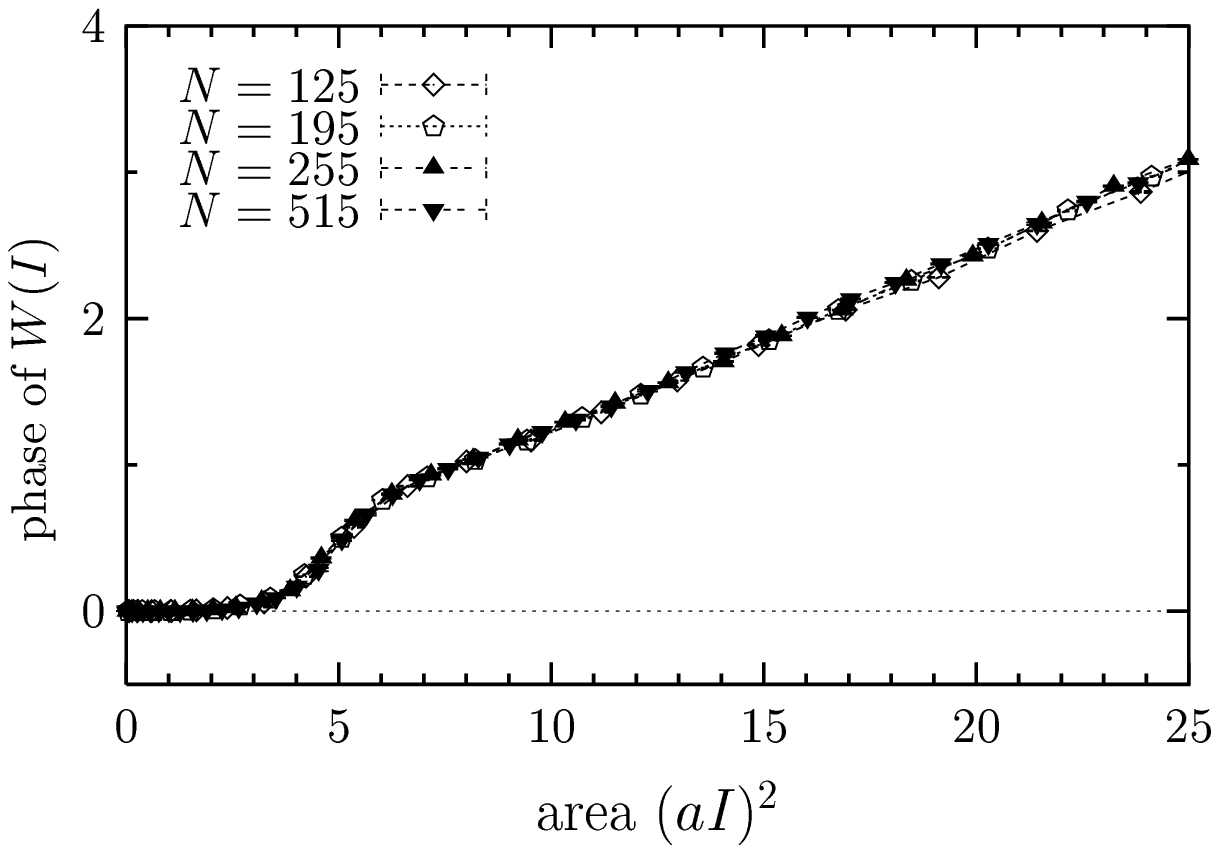}
 \end{center}
 \vspace{-2.5cm}
 \caption{{\it The polar coordinates of the complex Wilson loop $W(I)$
     plotted against the physical area $A = a^2 I^2$.
     At small areas it follows the Gross-Witten
     area law (solid line). 
     At larger areas the absolute value does not decay any more, 
     but the phase grows linearly.}}
 \label{wil_polar}
\end{figure}

\begin{figure}[htbp]
 \begin{center} 
   \includegraphics[width=.48\linewidth]{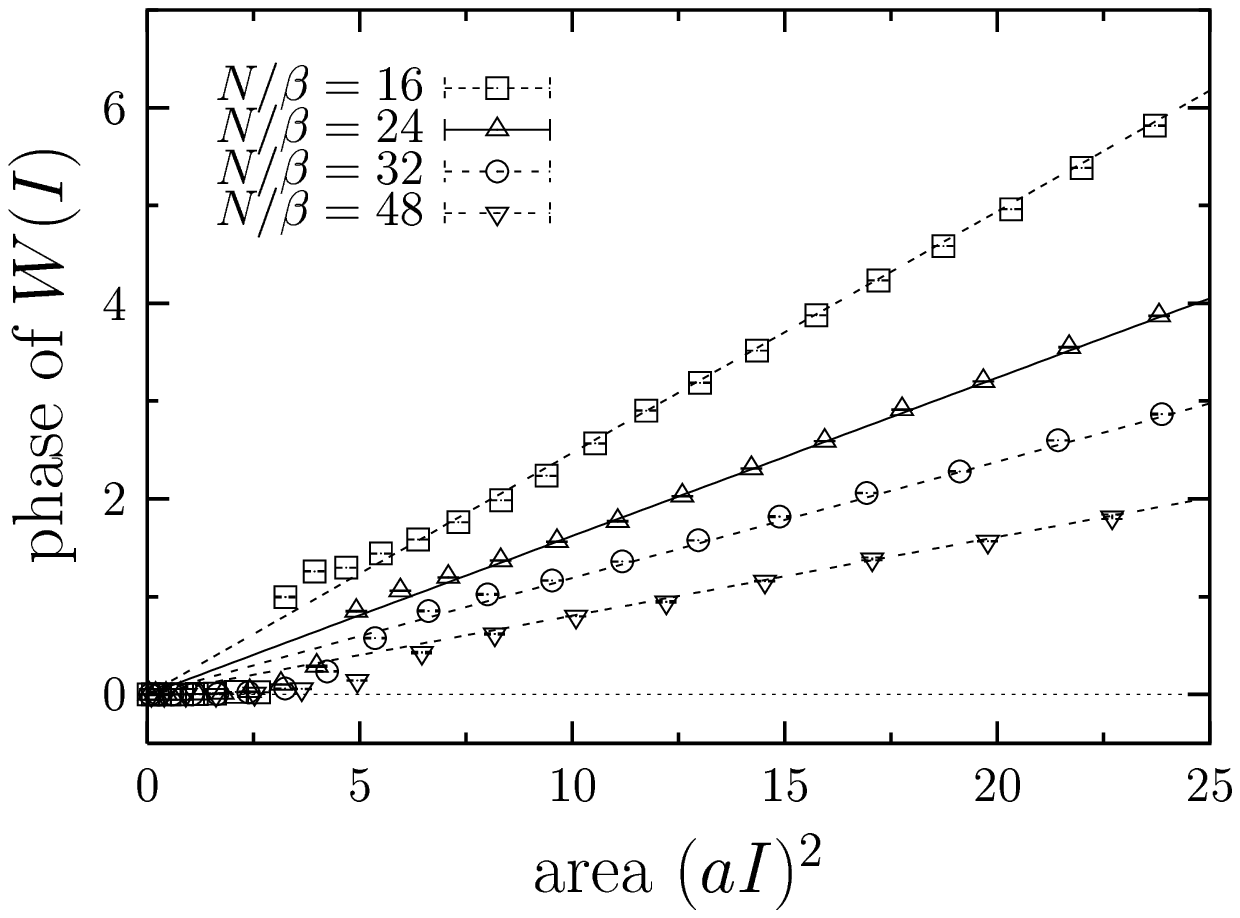}%
   \hspace*{.5cm}\includegraphics[width=.48\linewidth]{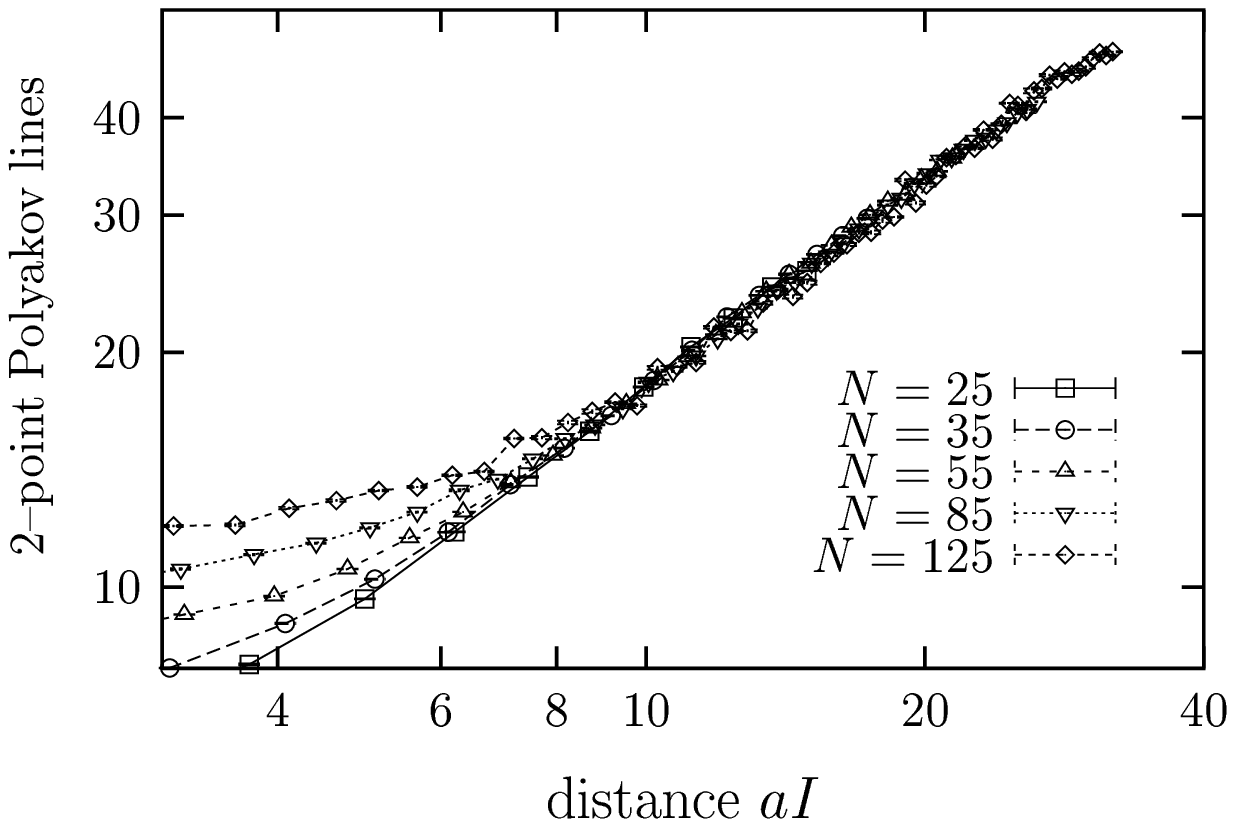}
 \end{center}
\vspace{-.5cm}
 \caption{{\it On the left: 
The phase of the Wilson loop $W(I)$ at $N=125$ for various
values of $N/\beta$, corresponding to different values of $\theta$.
At large areas $A >  O(\theta )$ the phase $\varphi$ agrees with 
the formula $\varphi = A/\theta$, which is shown by straight lines.
On the right:
The Polyakov 2-point function at $N=195$, after performing a universal
wave function renormalization, which provides a double scaling regime.}}
\label{2point}
\vspace*{-8mm}
\end{figure}

The (analogue of the) square shaped Wilson loop in the TEK takes the form
\beq
W_{12}(I\times I) = Z_{12}^{I^{2}} \, {\rm Tr}
(U_{\mu}^{I} U_{\nu}^{I}U_{\mu}^{\dagger \, I} U_{\nu}^{\dagger \, I}) 
= W_{21}^{*} (I \times I) \ ,
\eeq
and it corresponds to the Wilson loop in NC gauge theory.
Note that the expectation value
$W(I) := \frac{1}{N} \langle W_{12}(I\times I) \rangle $ 
is complex. Fig.\ \ref{wil_polar} shows its behavior in polar 
coordinates (for $N/\beta = 32$).
The large $N$ double scaling reveals that the Wilson loop is indeed
non-perturbatively renormalizable. At small area (relative to $\theta$)
the result follows the Gross-Witten area law, hence in this regime
the double scaling limit coincides with the planar limit of both,
the commutative and the NC model. At larger areas, however, 
$\vert W(I) \vert $ does not decay any further, but the phase $\varphi$
increases linearly in the area. Comparison of different values for
$N/\beta $ shows in that regime the simple relation
\beq 
\varphi = A / \theta = AB \qquad (A = (aI)^{2} = {\rm physical ~ area})
\eeq
to hold to a very high precision. Here $B$ is a magnetic field across
the plane, which we introduce formally as $B=1/\theta$, in agreement with
the Seiberg-Witten map \cite{SW99} and with solid state applications
\cite{condmat}. This relation is illustrated in Fig.\ \ref{2point} 
(on the left).

We also found large $N$ double scaling in some regime for the Wilson
2-point function and for the 2-point function of the Polyakov line
(see Fig.\ \ref{2point} on the right), if a suitable (universal) wave 
function renormalization is applied \cite{2dU1}.

\subsection{The NC $\lambda \phi^{4}$ model in three dimensions}

The NC $\lambda \phi^{4}$ model has been studied analytically
with various techniques. As a qualitatively new feature (compared
to the commutative world), it was conjectured that in part of the ordered
regime (at strongly negative $m^{2}$, which roughly corresponds to low
temperature) {\em stripe patterns} may dominate over the Ising-type uniform
ordering. Gubser and Sondhi performed a self-consistent Hartree-Fock
type one-loop calculation \cite{GS}, which would be exact in the $O(N)$ 
model at large $N$. They conjectured that it could also apply to the 
$\lambda\phi^{4}$ model, which suggests a striped phase in $d=4$ 
and $d=3$, but not in $d=2$ (in the dimensions $d<4$ they considered
an effective action as suggested by Brazovskii \cite{Bra}).
However, stripes did occur in 2d simulation results 
by Ambj\o rn and Catterall \cite{AmCat}; we add some remarks
on that case in the Appendix. In $d=3$ and 4, Gubser and Sondhi
predicted a first order phase transition from
the uniform to a striped phase if $\theta$ increases in the
ordered regime. At very strong $\theta$ they expect
more complicated patterns (like rectangles or rhombi) to become
stable.

Chen and Wu carried out a renormalization group analysis \cite{CW} in 
$d=4-\varepsilon$. They predicted a striped phase for 
$\theta > 12/\sqrt{\varepsilon}$, which should therefore not occur
in $d=4$.

\begin{figure}[htbp]
\vspace*{-1mm}
  \centering
  \includegraphics[width=.63\linewidth]{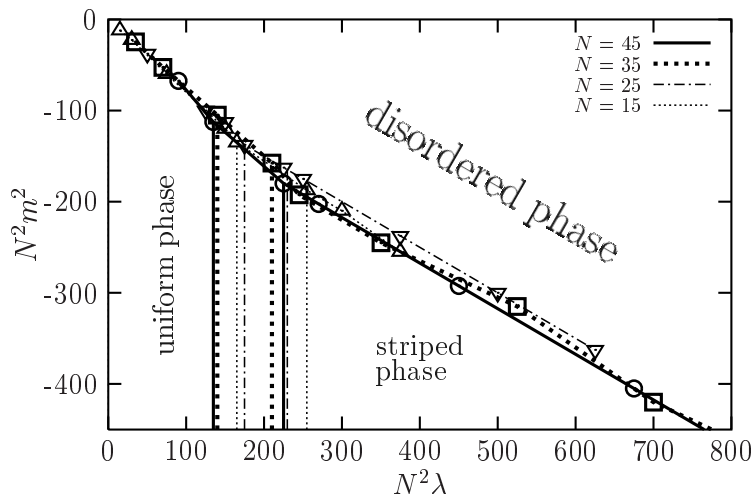}\\
\vspace{-2mm}
  \caption{\it The phase diagram of the 3d NC $\phi^4$ theory. 
    The connected points 
    show the separation line between the disordered and the ordered regime,
    and the vertical lines mark the transition region between the uniformly
    ordered and the striped phase.}
  \label{phasedia}
\end{figure}

\begin{figure}[htbp]
  \centering
\subfigure[{$N^2m^2=-22.5$}]{\epsfig{figure=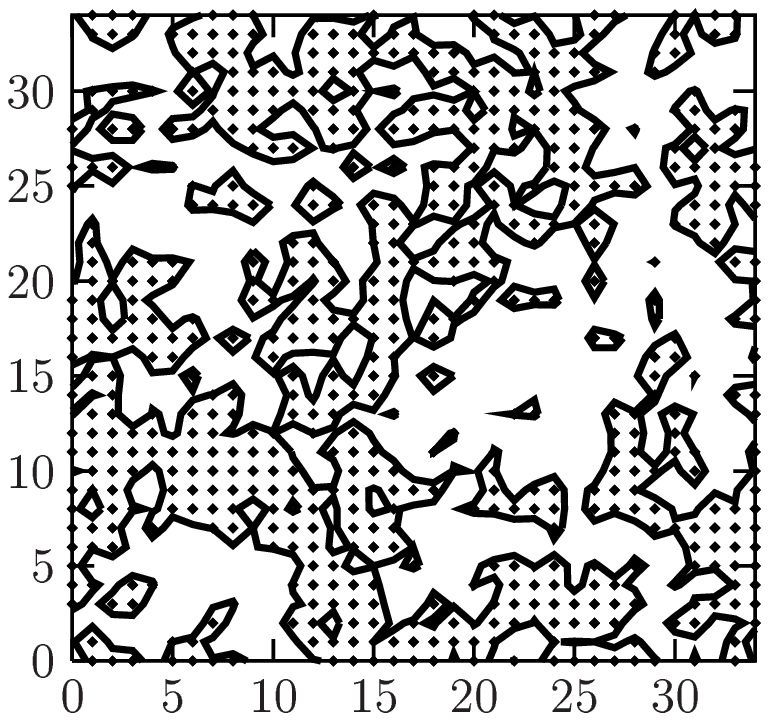,width=.2\linewidth}}%
\hspace{.5cm}\subfigure[{$N^2m^2=-360$}]
{\epsfig{figure=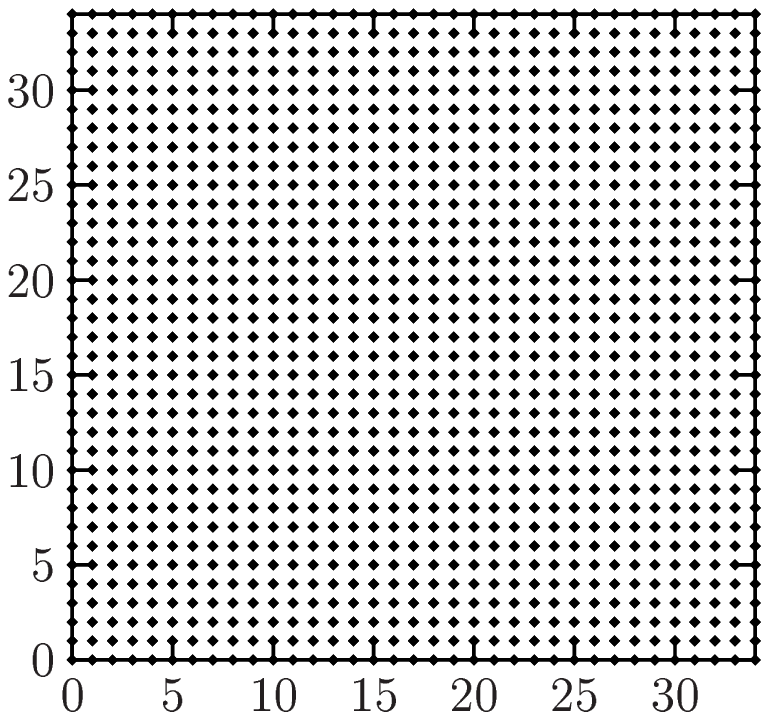,width=.2\linewidth}}%
  \hspace{1.5cm}\subfigure[$N^2m^2=-270$]
{\epsfig{figure=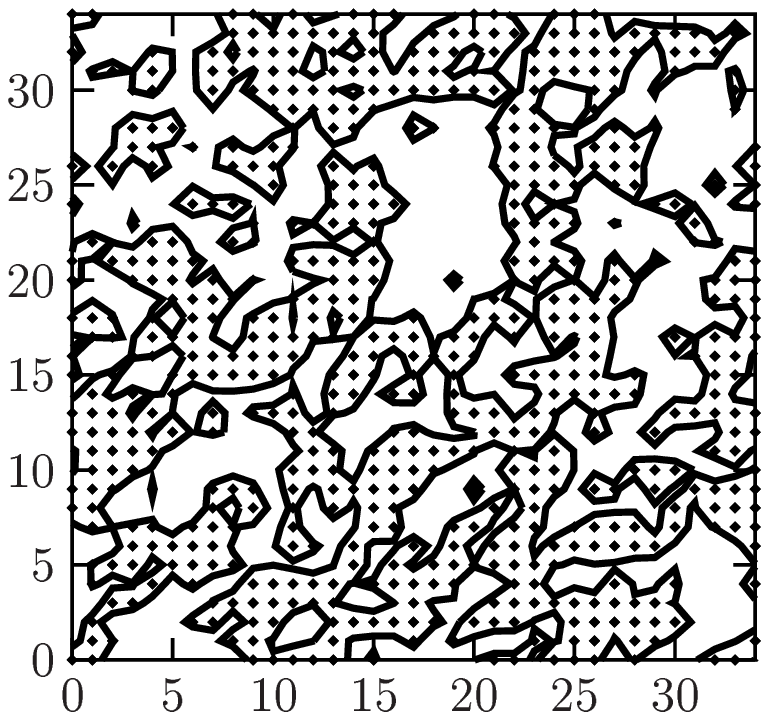,width=.2\linewidth}}%
  \hspace{.5cm}\subfigure[$N^2m^2\!=\!-1170$]
{\epsfig{figure=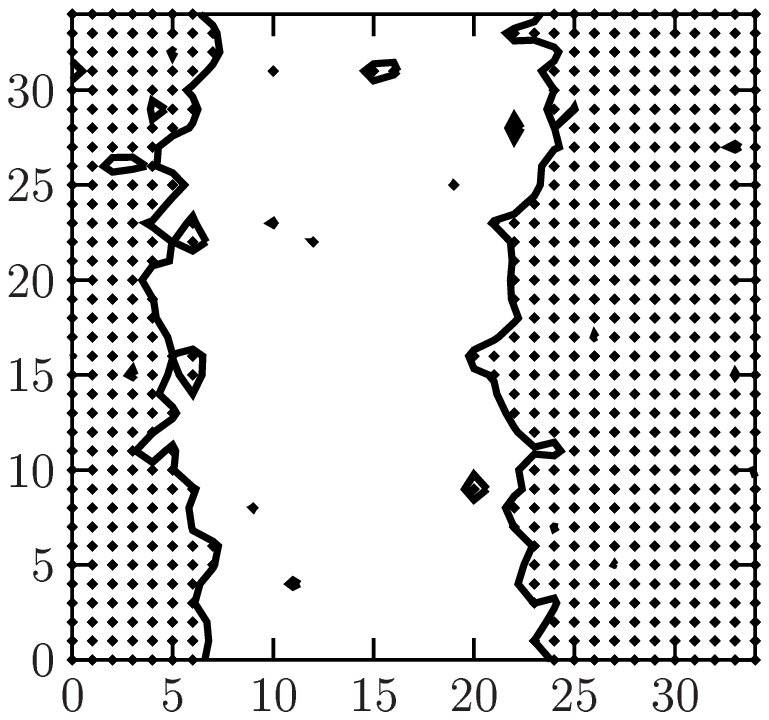,width=.2\linewidth}}
  \caption{\it Snapshots of typical configurations 
$\phi(t, \vec{x})$ at some time $t$, at $N=35$, 
$N^2\lambda=70$ (a,b; disordered,uniform) and $N^2\lambda=700$ 
(c,d; disordered,striped). We show the $(x_{1},x_{2})$ plane,
where dotted and blank regions correspond to the
different signs of $\phi$.}
  \label{snap}
\end{figure}

We considered this model in $d=3$ with two NC coordinates, and simulated
the corresponding matrix model (described in Section 1) at 
$T=N=15,\, 25,\, 35$ and 45.
Thus we obtained the explicit phase diagram shown in Fig.\ 
\ref{phasedia}. It is based on the order parameter
\begin{equation}
 \label{order-parameter}
  M(k)=\frac{1}{NT}\max_{|\vec{n}|=k}
\left|\sum_{t=1}^{T} \tilde{\phi}(t, \vec{p}) \right| \ ,
\qquad \vec p = \frac{2\pi}{N} \vec n \ , \quad
n_{1},n_{2} \in \{ 0,1,\dots ,N-1 \} \ .
\end{equation}
$\langle M(k) \rangle$ detects a uniform order at $k=0$ and a 
pattern of two stripes at $k=1$. Typical snapshots of ordered
and disordered configurations at small and at large $\lambda$
are shown in Fig.\ \ref{snap}.
We did not find more complicated patterns to be stable,
but we consider it as an open question if this may set in at larger
$N$ or $\lambda$ (we recall that this corresponds to larger $\theta$).
The order-disorder transition seems to be of second order, whereas
the uniform-striped transition appears to be first order --- it
takes the factor $N^{2}$ on the axes to stabilize it.
In order to localize the transitions accurately we also looked for
peaks in the connected 2-point functions of $M(k)$; some examples
were given in Ref.\ \cite{Lat01}.

\begin{figure}[htbp]
  \centering
  \subfigure[{$N^2m^2=-45$}]{\epsfig{figure=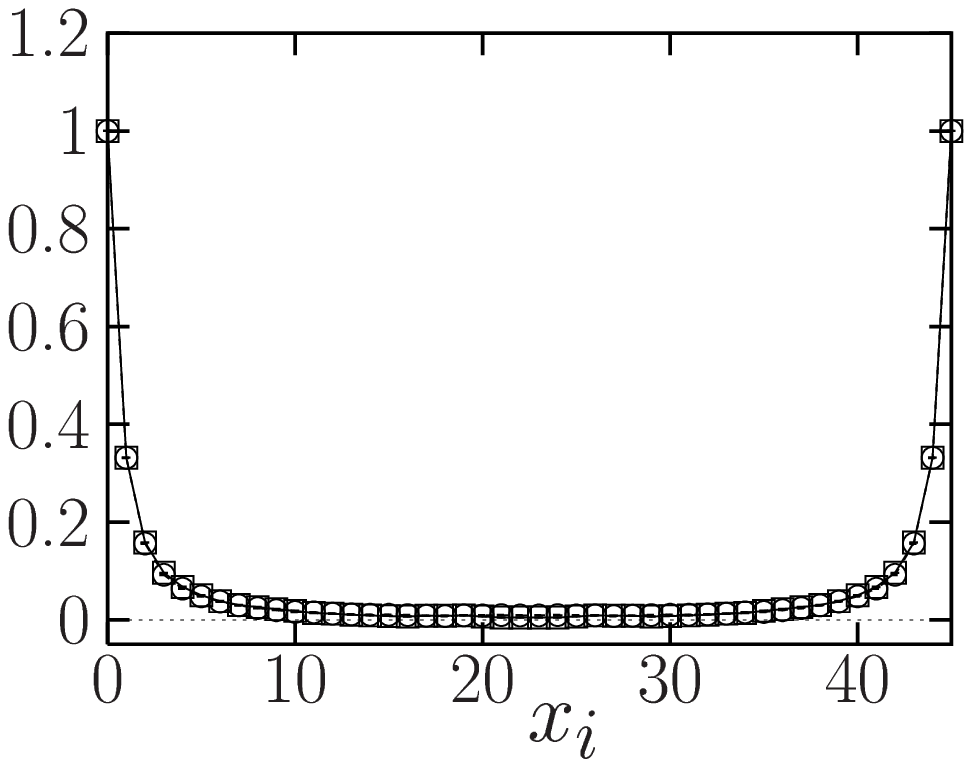,
width=.21\linewidth}}%
  \hspace{.5cm}\subfigure[{$N^2m^2=-225$}]
{\epsfig{figure=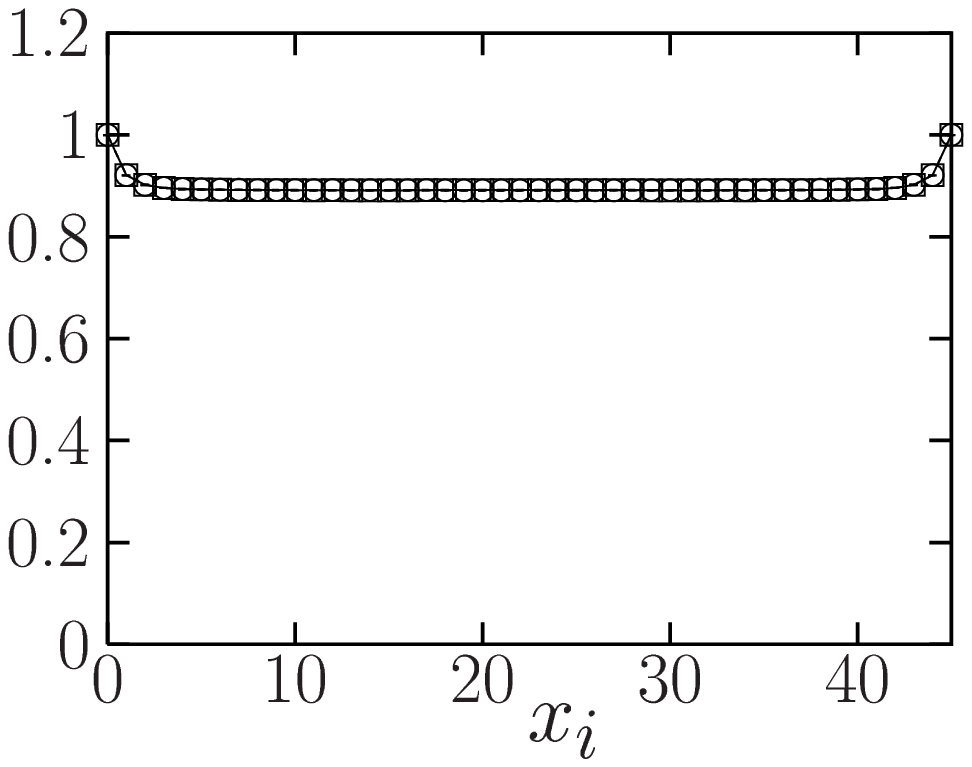,width=.21\linewidth}}%
  \hspace{1.2cm}\subfigure[$N^2m^2=-360$]
{\epsfig{figure=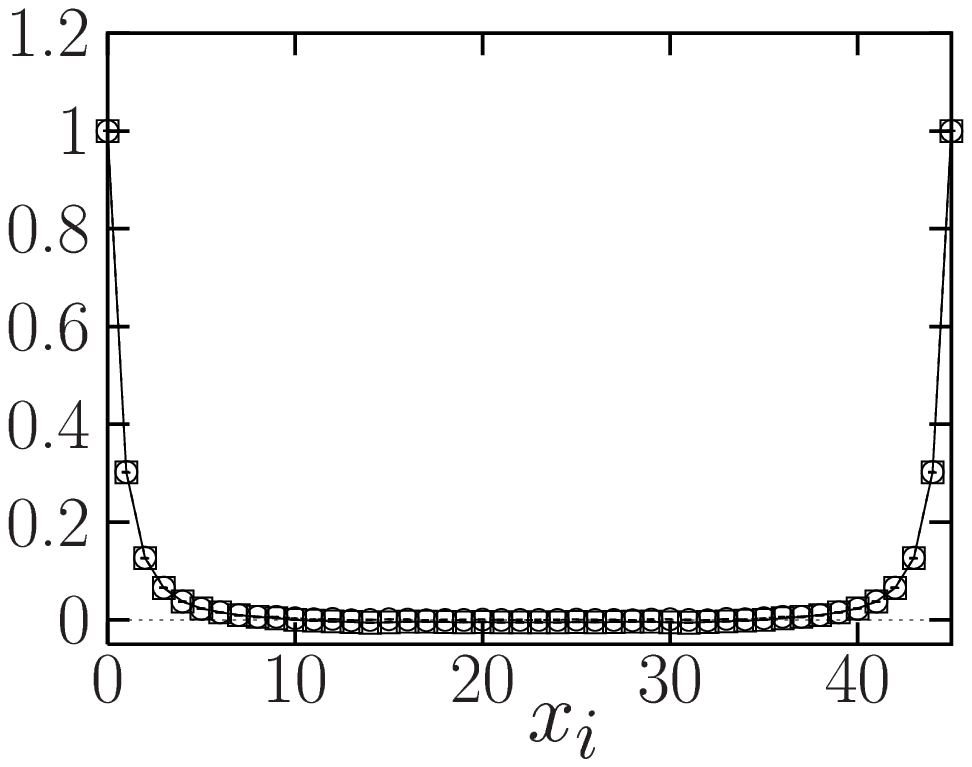,width=.21\linewidth}}%
  \hspace{.5cm}\subfigure[$N^2m^2=-945$]
{\epsfig{figure=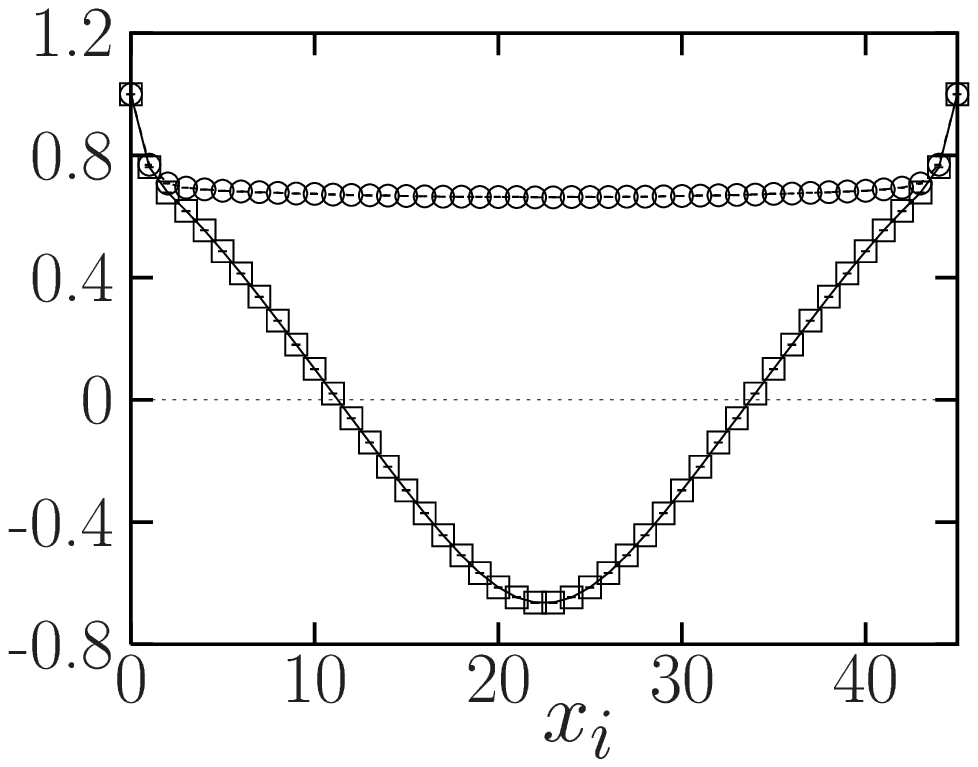,width=.21\linewidth}}
  \caption{\it The spatial correlator $C(\vec{x})$ 
defined in equation (\ref{space-corr})
    at $N^2\lambda=90$ (a,b) and $N^2\lambda=900$ (c,d) at $N=45$,
    against $x_i$. The circles and squares correspond to two orthogonal
directions; in particular in plot (d) the circles (squares) show the 
correlator parallel (vertical) to the stripes.}
  \label{spatial}

\end{figure}

\begin{figure}[htbp]
  \centering
  \includegraphics[width=.42\linewidth]{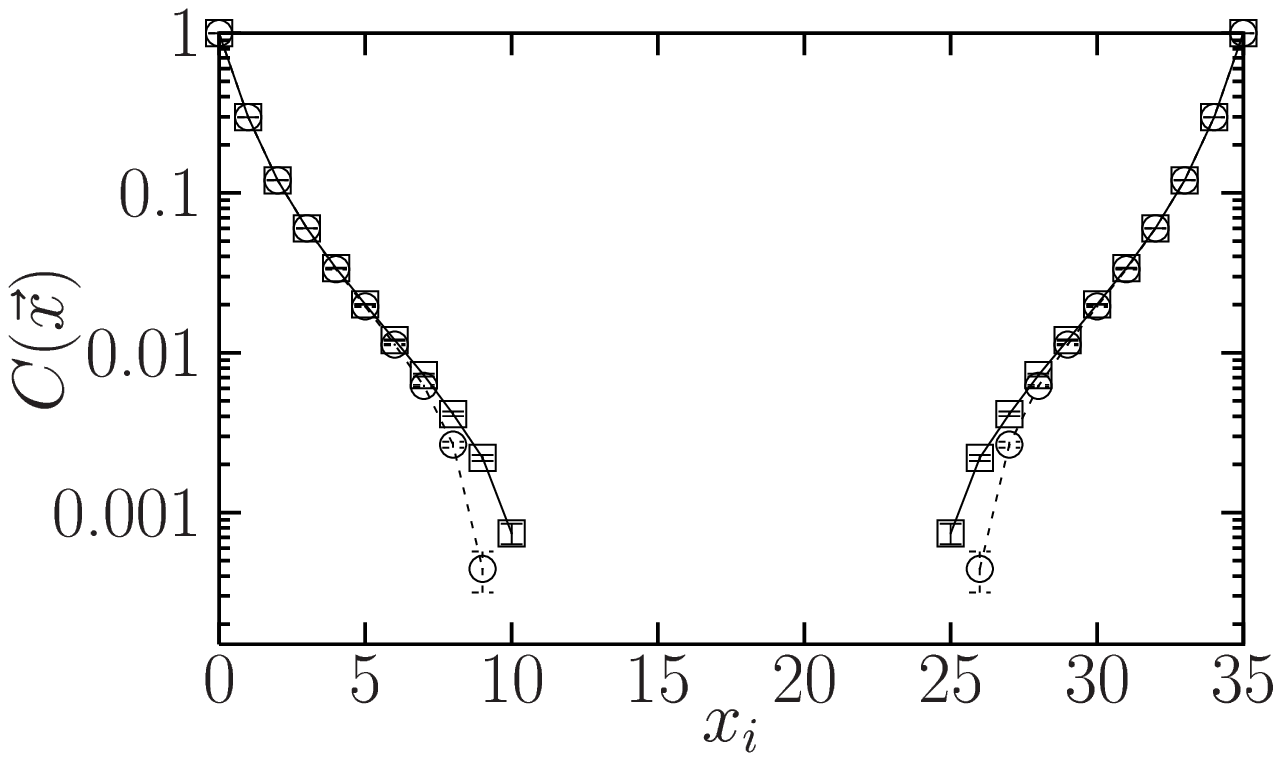}%
  \hspace{.7cm}\includegraphics[width=.42\linewidth]{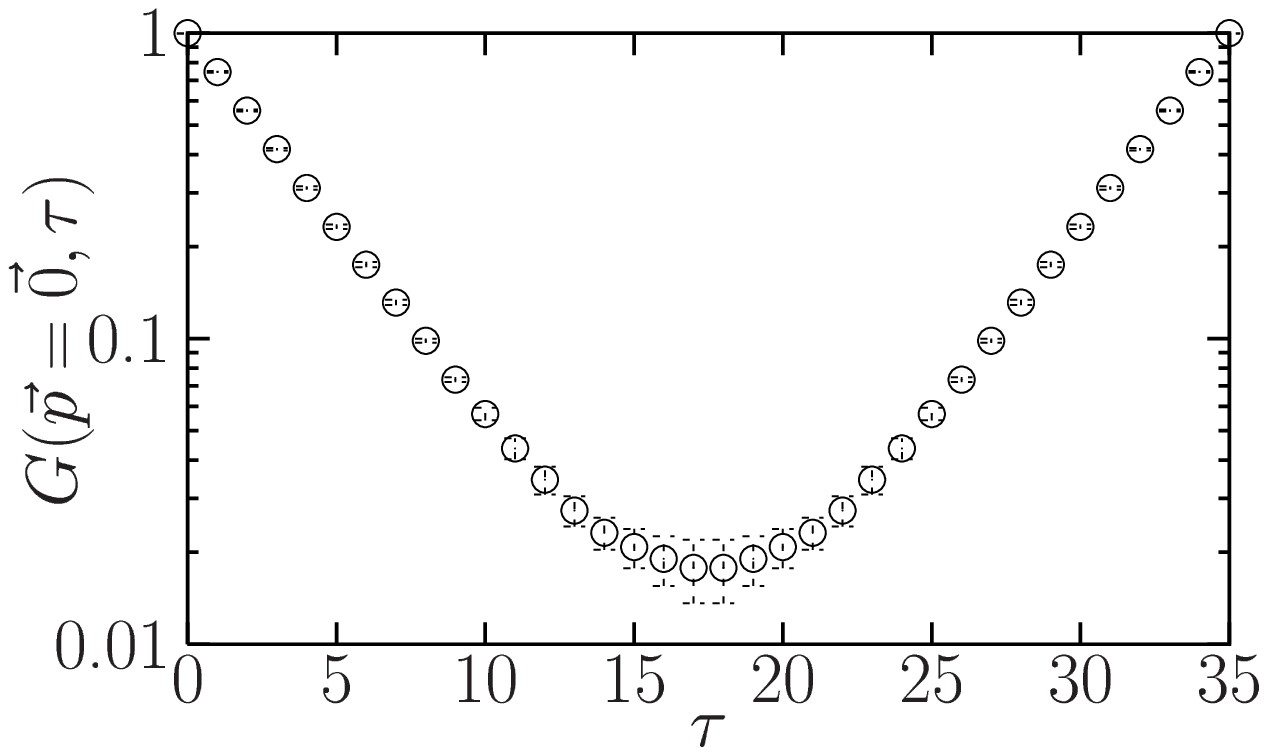}
\vspace*{-2mm}
  \caption{\it Correlation functions at $N=35$ in the disordered 
phase (near the phase transition) at $N^2m^2=-70$, $N^2\lambda=490$.
We show the spatial correlator (\ref{space-corr}) on the left,
and the temporal correlator in momentum space (\ref{time-corr}) at
$\vec{p}=\vec{0}$ on the right.}
  \label{temporal}
\end{figure}

Here we illustrate the different phases further by showing
the spatial correlator
\begin{equation}  \label{space-corr}
C(\vec{x}) = \frac{1}{TN^2} 
\sum_{t=1}^{T} \langle \phi(t, \vec{0})\phi(t, \vec{x}) \rangle
\end{equation}
in Fig.\ \ref{spatial}. It does not decay exponentially
in the disordered phase, see Fig.\ \ref{temporal} on the left.
Hence NC effects are visible also in the disordered phase.

\begin{figure}[htbp]
\vspace{-3mm}
  \centering
  \includegraphics[width=.42\linewidth]{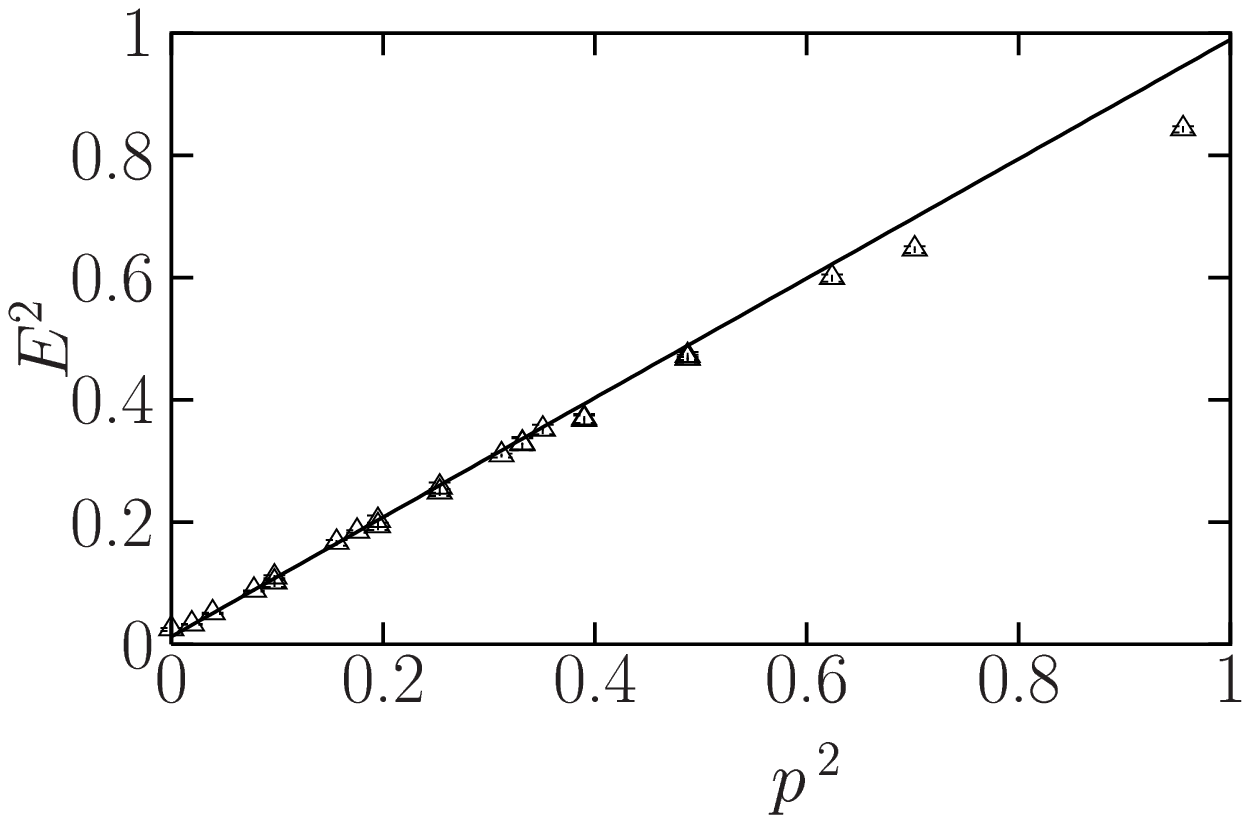}%
  \hspace{.7cm}\includegraphics[width=.42\linewidth]{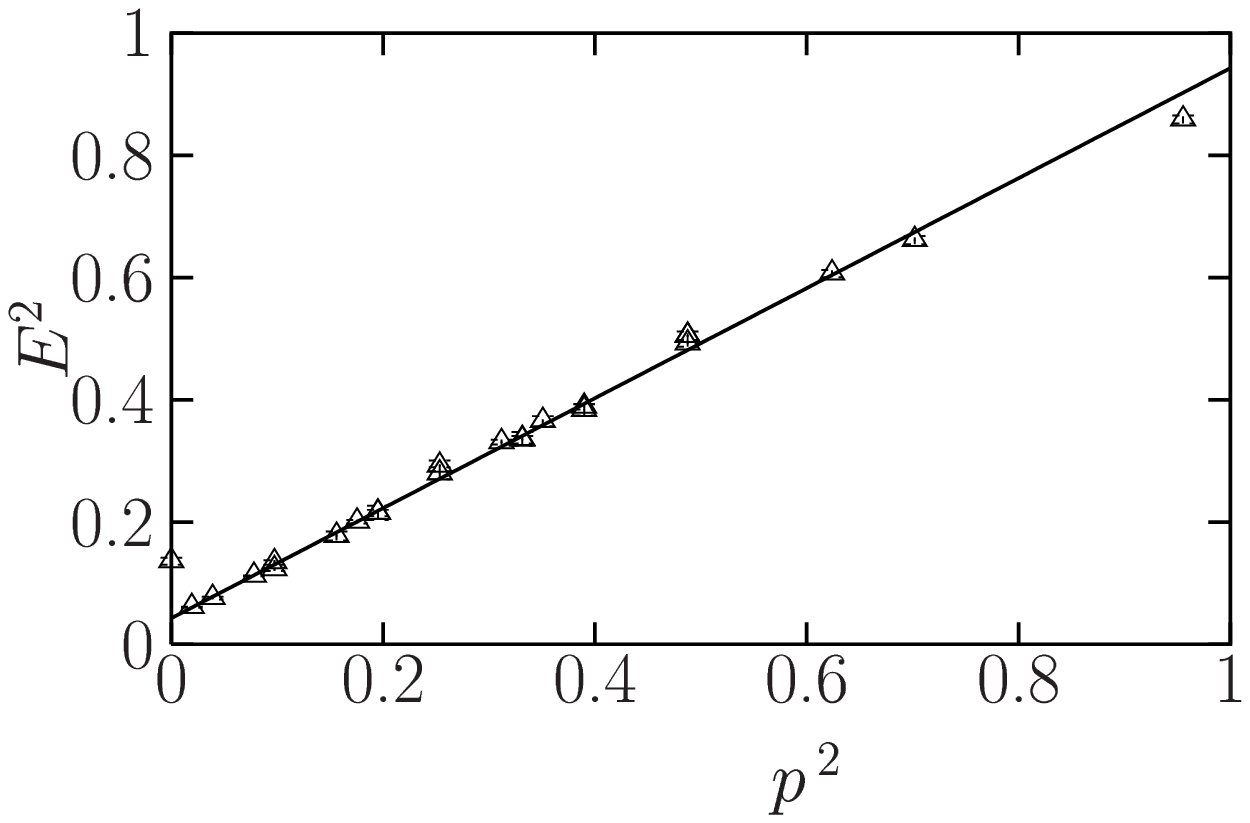}
\vspace*{-2mm}
  \caption{\it The dispersion relation $E^{2}( \vec p^{\, 2})$ 
in the disordered phase at $N=45$. We are close to the uniform
phase $N^2 m^2=-22.5,\, N^2 \lambda=90$ (left) resp.\ close to the
striped phase at $N^2m^2=-280,\,N^2 \lambda=900$ (right). The
solid line represents the continuum behavior in a Lorentz invariant
theory.}
  \label{disp}
\vspace*{-3mm}
\end{figure}

On the other hand, the temporal correlator in momentum space
\begin{equation}  \label{time-corr}
  G(\tau , \vec{p})= \frac{1}{TN^2}\sum_{t=1}^{T} {\rm Re}
  \left\langle\tilde{\phi}(t,\vec{p})^*\tilde{\phi}
(t+\tau ,\vec{p})\right\rangle
\end{equation}
does follow an exponential decay (resp.\ a {\tt cosh} behavior at 
finite $T$). Fig.\ \ref{temporal} on the right is an example
at $\vec p = \vec 0$. Varying $\vec p$ we can identify the dispersion
relation $E(p)$, which we show in Fig.\ \ref{disp}, still in the 
disordered phase. It follows closely the linear behavior 
$E^{2} = \vec p^{\, 2} + M_{\rm eff}^{2}$ (solid line, where
$M_{\rm eff}$ is a free parameter). 
The deviation at large momenta is a
trivial lattice artifact. However, we also observe a systematic jump at
$\vec p = \vec 0$, only in the case where we are close to the
striped phase. This confirms the energy minimum at 
$k = N\vert \vec p \vert /(2\pi ) =1$, i.e.\ the trend towards stripes
rather than uniform ordering.
Such irregularities in the dispersion at significant
$\theta$ may be of importance; note that a $\theta$ deformed 
dispersion relation has also been postulated for photons \cite{photon}.


\section{Conclusions}
\vspace{-1mm}

We presented non-perturbative results for two types of NC field
theories. Such theories are non-local in the range of 
$O(\sqrt{\theta })$. We observe basic deviations from the
commutative world rather at large scales though, which can be 
understood from the UV/IR mixing.

Simulating the TEK we obtained results for 2d NC $U(1)$ gauge theory.
The scaling behavior that we observed is the first evidence for the
non-perturbative renormalizability of a NC field theory.
The Wilson loop follows an area law at small area. At large area
it can be described by the Aharonov-Bohm effect if we identify
$B = 1/\theta $, which appears here as a dynamical relation
(on tree level it was used before in the literature).

In the 3d $\lambda \phi^{4}$ model on the lattice we explored the
phase diagram in the $m^{2}$ -- $\lambda$ plane. In the ordered regime
(at strongly negative $m^{2}$) we found at small $\lambda $ a uniform
order of the Ising type, as in the commutative case. At larger
$\lambda$ --- which amplifies the NC effects --- we observed a pattern of
two stripes to be stable for $N=15 \dots 45$. It is an open question
if multiple stripes etc.\ stabilize at larger values of $N$ or $\lambda$, 
as conjectured by Gubser and Sondhi. 
We observed the same behavior in $d=2$, where we
also verified up to $N=45$ that only two stripes are ultimately
stable (see Appendix).
Still this implies the spontaneous breaking of translation
symmetry, which is possible due to the non-locality \cite{AmCat}.

In the ordered regime, the spatial correlations
are dictated by the dominant pattern: uniform as in the
commutative case, or striped with strong correlation in the direction
of the stripes and anti--correlation vertical to them.
\footnote{The stripes that we observed were always parallel to 
one of the axes.}
This agrees with the predictions in Ref.\ \cite{GS}.
In the disordered phase the spatial correlators deviate from
the exponential decay. However, the correlators in momentum space do
decay exponentially in time for all momenta.

That property allowed us to study the dispersion relation
in the disordered phase: at small $\lambda$ also the dispersion
behaves qualitatively as in the commutative model, but at large
$\lambda$ there appears a jump at $\vec p = \vec 0$ as a NC effect.
Hence the energy minimum is at $\vert \vec p \vert =2\pi /N$, 
in agreement with
the trend towards two stripes. It remains to be clarified how this
irregular dispersion behaves at large $N$, and if it can be related
to the expected non-linear photon dispersion at $\theta >0$; the latter
makes experimentalists hope for a precise measurement of $\theta$
(for a discussion, see for instance Ref.\ \cite{astro}).\\

{\em Acknowledgement} \ 
We thank J.\ Ambj\o rn, A.\ Barresi, S.\ Catterall, L.\ Doplicher,
H.\ Dorn, K.\ Fredenhagen, D.\ L\"{u}st, 
X.\ Martin, D.\ O'Connor, P.\ Pres\v{n}ajder, R.\ Szabo 
and P.\ van Baal for illuminating discussions.

\appendix 

\section{The NC $\lambda \phi^{4}$ model in $d=2$}

The occurrence of stripes in the ground state implies the
spontaneous breaking of translation invariance. The case
$d=2$ is particularly interesting in this respect. Gubser and
Sondhi argued based on the Mermin-Wagner Theorem that stripes
cannot be stable in $d=2$. However, Ambj\o rn and Catterall
pointed out that this Theorem is in general not applicable
to non-local theories, and in fact they did observe non-uniform
patterns in their numeric results for the NC $\lambda \phi^{4}$
model in $d=2$ \cite{AmCat} (where the two coordinates obey the 
NC relation (\ref{standardNC})).
\footnote{One might expect that the Theorem extends to theories, which 
are non-local in only a small range, which is invisible at very
large scales. However, this argument is not adequate for NC models
due to the UV/IR mixing.}

\begin{figure}[htbp]
  \centering
  \subfigure{\epsfig{figure=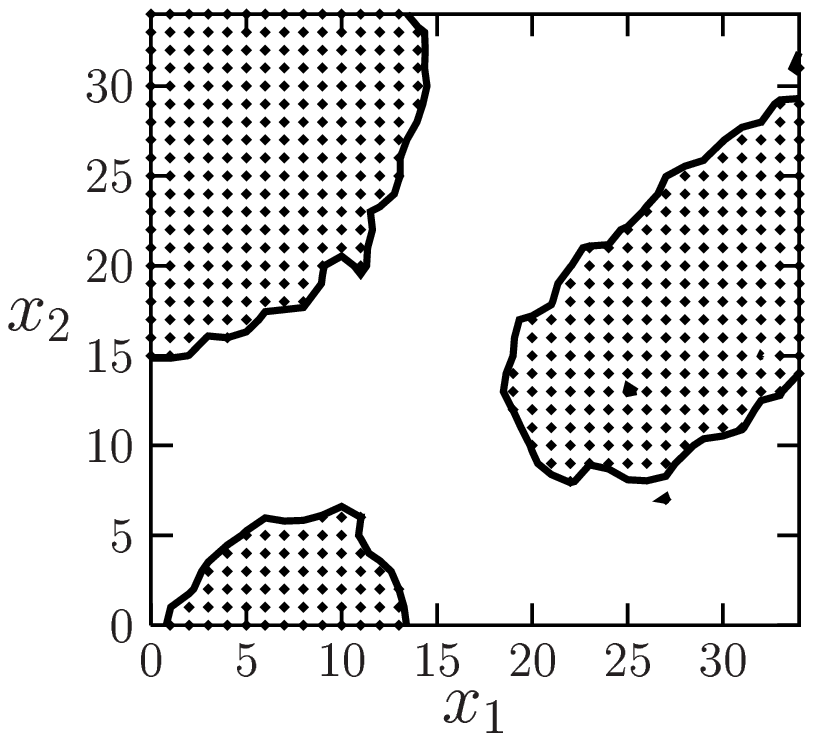,width=.2\linewidth}}%
  \hspace{.5cm}\subfigure{\epsfig{figure=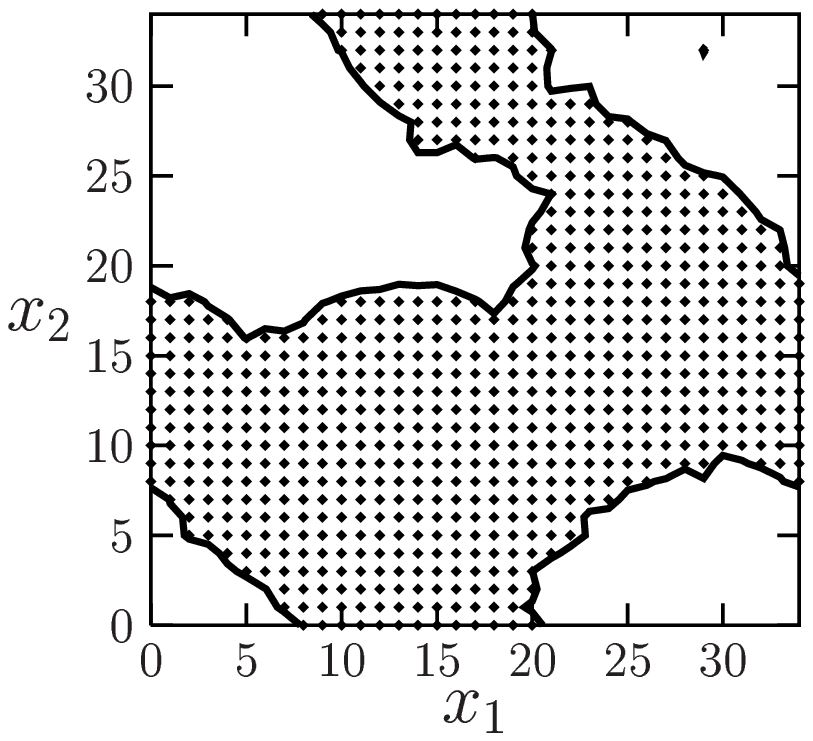,width=.2\linewidth}}%
  \hspace{.5cm}\subfigure{\epsfig{figure=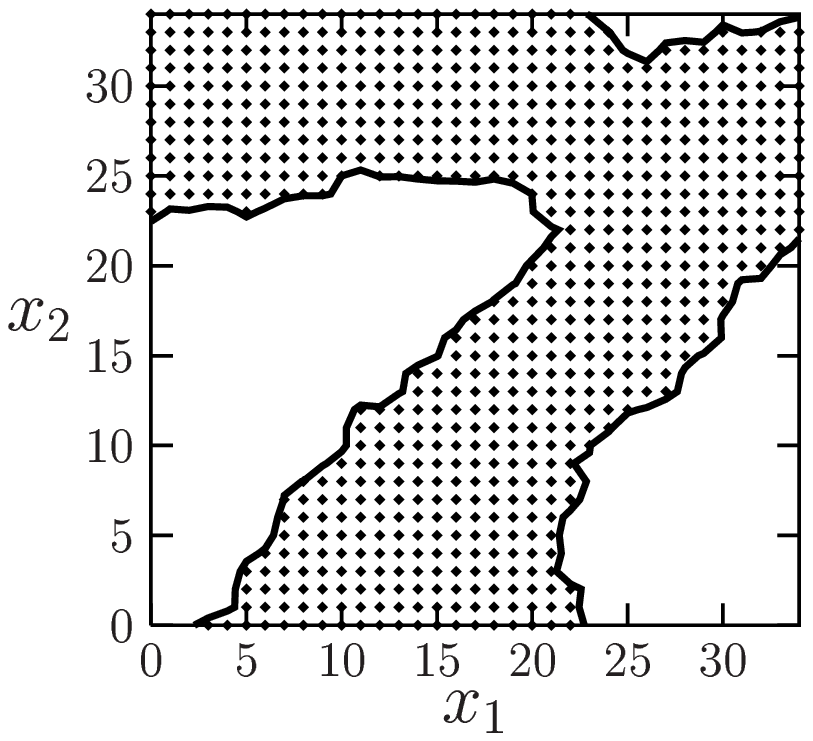,width=.2\linewidth}}%
  \hspace{.5cm}\subfigure{\epsfig{figure=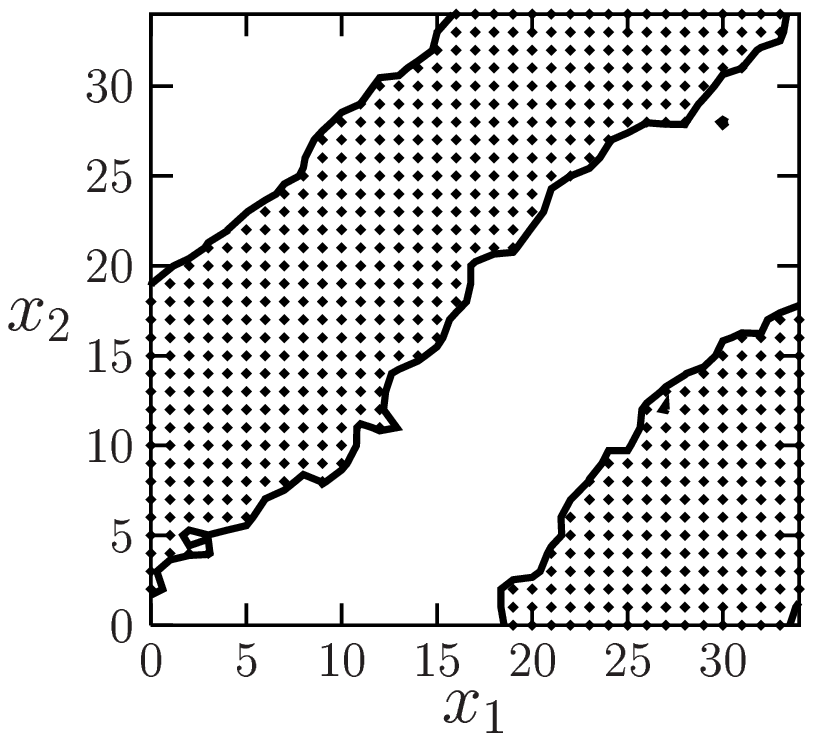,width=.2\linewidth}}
\vspace{-3mm}
  \caption{\it Examples for meta--stable patterns in two dimensions at  
$N=35$, $\lambda=0.43$, $m^2=-2$, which is deeply in the striped phase.
We show the same sort of maps as in Fig.\ \ref{snap}.}
   \label{metastab-2d}
\vspace*{-3mm}
 \end{figure}

We also performed simulations in $d=2$, and we fully agree that also
there stripe vacua exist. Starting from hot configurations
and proceeding in Metropolis steps which propose a full Hermitian matrix 
to be added to $\hat \phi (t)$ (with a coefficient tuned for a good
acceptance rate), we find after about 500 steps a rich structure of
patterns, where in general $p_{1}$ and $p_{2}$ are both non-zero, and 
often larger than $2\pi /N$. However, if we continue to thermalize much 
longer we see that only the two-stripe pattern is ultimately stable,
just as in $d=3$. 
\footnote{However, we verified carefully, both in $d=2$ and in $d=3$,
that those two stripes are really stable; one consistently runs into them
starting from all sort of hot configurations, and they show absolutely no
trend to disappear in very long histories.}
This we tested again up to $N=45$.
The same can be seen if we switch to a more efficient algorithm, which 
updates the (conjugate pairs of) matrix elements one by one (this
reduces the thermalization time by a factor $\propto 1/N$).
Still the variety of meta-stable patterns may be of importance;
we present some examples for them in Fig.\ \ref{metastab-2d}.
The Goldstone Theorem still holds for NC field theories \cite{NCGB},
hence it is an interesting question to investigate the massless
fluctuations related to translation symmetry.
 
\end{document}